\documentstyle[PASJadd]{PASJ95}
\draft
\begin{document}

\title{Usability of the NFW Galaxy Profile as a Cosmological Tool
Estimated from 2-Image Gravitational Lens Systems}
\author{Katsuaki {\sc Asano} \\
{\it Department of Physics, Ritsumeikan University,
	Kusatsu, Shiga 525-8577} \\
{\it E-mail: sph10001@se.ritsumei.ac.jp}}

\abst{
The profile of Navarro, Frenk, and White (the NFW profile),
which was derived from the {\it N}-body simulations of cold dark-matter halos,
is a strong candidate for a galaxy or cluster profile.
In order to check the usability of the NFW profile as
a first approximation of a galaxy model,
we studied the characteristic overdensity and scale radius of galaxies
by reproducing the image positions and flux ratios
of 2-image gravitational lens systems,
under the following simple assumptions: the galaxies are spherically symmetric,
and stars and external shear do not contribute to the gravitational lens.
The scale radii of the lensing galaxies are smaller, and the characteristic overdensities
are larger than the predicted value in the {\it N}-body simulations.
These results indicate that our assumptions are overly simplified.
It may be impossible to simply adopt the NFW profile, which does not include stars,
to probe the cosmological parameters or the light propagation in
an inhomogeneous universe and so on.
If we adopt a softened isothermal profile to the lensing galaxies,
the scale radii and the central matter densities agree with
models which are derived from other observational results for early-type galaxies
and which are independent of gravitational lensing.
The isothermal sphere as a first approximation of a galaxy model
has no serious problem.
}
\kword{dark matter---galaxies: structure---gravitational lensing}

\maketitle
\thispagestyle{headings}


\newpage

\section{Introduction}

\indent

Gravitational-lens systems provide a powerful tool to probe the potentials of galaxies,
the cosmological parameters, light propagation in an inhomogeneous
universe and so on.
In an analysis of gravitational-lens systems or other observations,
we need to assume a profile
of the mass distribution of galaxies.
The softened isothermal profile and the modified King
profile have been conventionally used to determine the Hubble constant, $H_{0}$,
or the density profiles of lensing galaxies (Kent, Falco 1988; Grogin, Narayan 1996;
Keeton, Kochanek 1997).
These profiles are approximately derived from the collisionless Boltzmann equation,
and explain the H {\small I} rotation curves well (Binney, Tremaine 1987).
If these profiles, however, do not represent the actual features of galaxies,
the cosmological parameters obtained need to be altered.
Thus, it is important that we correctly select the mass profiles of galaxies.

There is no definite evidence that an isothermal profile is not appropriate.
We should, however, compare the isothermal profile with other plausible profiles.
One of the most highly possible candidates for galaxy profiles,
instead of the isothermal profile,
is the profile described by Navarro, Frenk and White (1995).
Using {\it N}-body simulations of
cold dark-matter halos,
Navarro, Frenk and White (1996, hereafter NFW96; 1997, hereafter NFW97)
revealed that a density cusp forms at the center,
and that the density profiles have
the same shape, independent of the halo mass and of the initial density
fluctuation spectrum. According to Navarro et al., the central density profile
fits the analytic form,
\begin{equation}
\rho(r)=\frac{\delta_{\rm c} \rho_{\rm cr} }{(r/r_{\rm s})(1+r/r_{\rm s})^2},
\label{nfw}
\end{equation}
(hereafter we refer to this as the NFW profile),
where $r_{\rm s}$, $\rho_{\rm cr}=3 H_0^2 /8 \pi G$, and $\delta_{\rm c}$
are the scale radius, the critical density and the characteristic
overdensity, respectively.
The NFW profile is found in {\it N}-body/hydrodynamic
simulations (Navarro et al. 1995; Eke et al. 1997).
They studied objects with masses ranging from those of
dwarf galaxy halos to those of rich galaxy clusters.
Carlberg et al. (1997) found from measurements
of galaxy velocity dispersions in clusters that the NFW profile
can reproduce the measured velocity dispersion.

The results for the scale radius and the characteristic overdensity
are summarized in table 1 in NFW96.
Navarro et al. predict that the scale radius is
approximately several hundred kpc at the cluster mass scale.
On the other hand, X-ray observations of galaxy clusters
do not agree with these results (Makino et al. 1998;
Makino, Asano 1999).
The size of $r_{\rm s}$ for galaxy clusters
predicted in NFW96 is smaller than the observed values.
The relation between the scale radius and the characteristic overdensity
of galaxies also needs to be probed.

In general, a large number of stars or gas should be incorporated
when we obtain accurate parameters of lens galaxies,
although the NFW profile is obtained from cold dark-matter simulations.
Complex models with many parameters
can greatly reproduce the observations of gravitational lenses.
However, it is not absolutely clear that
the rough parameter values of the dark halo can be
obtained from the NFW profile, which does not include stars or gas.
If we can use the NFW dark halo as a first approximation of a galaxy model,
the NFW profile would be very convenient to apply it to various cosmological analyses.
On the other hand,
the isothermal profile has been solely adopted as a first approximation.
The light propagation in an inhomogeneous universe have studied
by applying an isothermal profile to galaxies or galaxy clusters
(e.g. Jaroszy\'nski 1992; Premadi et al. 1998, Tomita et al. 1999).
In this kind of analysis, we need simple models of the galaxy profile,
while a complex model with many parameters and perturbations
is needed to obtain the Hubble constant from gravitational lenses.
The parameter values of the NFW profile has been predicted by numerical simulations.
If we can neglect the contribution of stars,
light propagation in an universe with galaxies of the NFW profile
can be promptly studied.
An analysis with the NFW profile may bring us different results from
those at present for light propagation.
For the above reasons, the usability of the NFW profile, which does not include stars
as a first approximation of the galaxy profile, should be checked,
as is the case with the isothermal sphere.
The usability of the NFW galaxy halo as a cosmological tool is not necessarily well-known.
In an analysis of galaxy clusters with the X-ray surface brightness
(Makino et al. 1998), the gravity of the gas is not very important
in order to obtain the density profile.
Of course, this situation is not the same for galaxies,
since their mass-to-light ratio is much smaller than those of galaxy clusters.
We expect that the mass-to-light ratio in the central regions of galaxies
is about 10 (e.g. Lauer 1985).
Sofue (1999) claimed that the bulge
of LMC has a large fraction of dark matter
(the mass-to-light ratio is about 20--50).
For spiral galaxies, a high-density concentration of dark mass was reported
(Sofue et al. 1997, 1998).
Thus, it is not necessarily obvious that the stellar component
predominates in the centeral region of galaxies
sufficiently enough to intervene the simple analysis of the dark halo profile.
We should confirm whether the NFW halo, without stars,
can be a rough approximation of galaxy profile or not.

In this paper we consider the characteristic overdensity
and the scale radius of galaxies for the NFW profile
from 2-image gravitational lens systems
and a simple lens model without stellar component.
The softened isothermal profile is also adopted for a comparison.
This profile represents the other conventionally used profiles.
The reason for studying this subject is
to confirm whether the NFW profile, which does not include stars or gas,
can be a cosmological tool to a first approximation.
We discuss whether the NFW profile, which is obtained from our method,
can express the actual order of galaxy parameters or not.
Our goal in the future is to check the consistency between
observations and the parameters predicted from the {\it N}-body simulations.
This study is the first step toward this goal.
Only the predicted parameter values in NFW96 are constraints for the
parameters of the NFW profile at present.
Bartelmann (1996) and Evans and Wilkinson (1998)
adopted a scale radius which satisfies an estimation by
numerical simulations of NFW96 for galaxy clusters.
Thus, our strategy in this paper is that we accept the results in NFW96
to be as Bartelmann et al. assumed,
and to compare the parameters obtained from the gravitational lenses
with those in NFW96.
The disagreement between NFW96 and the results of gravitational lenses
indicates that the galaxy models which do not include stars are not adequate,
or that the results in NFW96 are not reliable.
If the results in NFW96 are wrong due to softening and small numbers of particles,
it also means that we cannot use the NFW profile for galaxies as a first
approximation at present.
Anyhow, we can clarify the usability of the NFW profile from this study.
We apply simple lens models, with no contribution of stars, no external shear,
and spherically symmetric lensing, as a first approximation.
Our aim is not to obtain accurate parameters of galaxies
by a precision analysis.
The usability of the NFW dark halo as a first approximation is our theme in this article.
In section 2 we explain the method of analysis.
We only consider spherical lenses so as to avoid complex assumptions.
2-image lens systems are good targets for this method (Schneider et al. 1992).
In section 3, our method is
applied to the observation of lens systems.
The results for the NFW and the softened isothermal profiles
are summarized.
The results for the softened isothermal profile are compared with
the galaxy models of Jaroszy\'nski (1992).
In section 4, we summarize our work, and discuss our results.

Throughout this paper, we assume $H_0=50~{\rm
km\,s^{-1}\, Mpc^{-1}}$, $\Omega_0=1$, and $\Lambda_0=0$.

\section{Method of analysis}

\indent

We consider 2-image lens systems in order to estimate the relation
between the galaxy parameters.
Since 2-image lens systems are approximately expressed by a
spherically symmetric lens (Schneider et al. 1992),
we can study the structure of lensing galaxies by the following
simple process without using the $\chi^2$ minimization technique.
We do not need to consider any complicated assumptions such as ellipticity,
external shear, or spiral structure.

The spherically symmetric lenses are completely described by
their surface mass density, $\Sigma(r)$,
and the critical surface mass density,
\begin{equation}
\Sigma_{\rm cr}=\frac{c^2}{4 \pi G} \frac{D_{\rm s}}{D_{\rm l} D_{\rm ls}},
\end{equation}
where $D_{\rm l}$ is the angular diameter distance to the lens,
$D_{\rm s}$ is that of the source,
and $D_{\rm ls}$ is the distance from the lens to the source.
By symmetry, the lens position, the source position, $y$, and
the image positions, $r$, are on one straight line.
By placing the lens at the origin,
the lens equation reduces
to a one-dimensional form, and is expressed by
\begin{equation}
y=r-\frac{m(r)}{r},
\label{lsq}
\end{equation}
where $m(r)$ is defined by
\begin{equation}
m(r)=2 \int_{0}^{r} \frac{\Sigma(r')}{\Sigma_{\rm cr}} r' dr',
\end{equation}
(Schneider et al. 1992).
For the NFW profile, one finds
\begin{equation}
m(r)=r_{\rm s}^2 \kappa_0 g(x),
\label{mr}
\end{equation}
where
\begin{equation}
g(x)=2 \ln{\frac{x}{2}}+
\left\{ \begin{array}{ll}
\displaystyle{\frac{4}{\sqrt{x^2-1}}} \tan^{-1}{\displaystyle{\sqrt{\frac{x-1}{x+1}}}} &
(x > 1) \\
1 & (x=1) \\
\displaystyle{\frac{4}{\sqrt{1-x^2}}} \tanh^{-1}{\displaystyle{\sqrt{\frac{1-x}{1+x}}}} &
(0 < x < 1), 
\end{array} \right.
\end{equation}
and $x=r/r_{\rm s}$ (Bartelmann 1996; Maoz et al. 1997).
The dimensionless parameter $\kappa_0$ is written by
\begin{equation}
\kappa_0=2 \delta_{\rm c} \rho_{\rm cr} r_{\rm s}/\Sigma_{\rm cr}.
\end{equation}
Here, we neglect the contributions of stars and gas as a first approximation.
We also consider the softened isothermal profile,
\begin{equation}
\rho(r)=\frac{\delta_{\rm c} \rho_{\rm cr} }{1+(r/r_{\rm s})^2}.
\end{equation}
Here, we used the same notation for the parameters as those for the NFW profile,
in order to simplify our formulation.
We should note that there is no direct relation between the parameters for
the NFW and the softened isothermal profiles.
In this notation, $m(r)$ for the softened isothermal profile is the same as equation
(\ref{mr}) with
\begin{equation}
g(x)=\pi (\sqrt{1+x^2}-1).
\end{equation}
In spherical lens systems,
two images are positioned on opposite sides of the lensing galaxy.
The observed image positions $r_{\rm A}>0$ and $r_{\rm B}<0$ should satisfy
\begin{equation}
r_{\rm A}-\frac{m(r_{\rm A})}{r_{\rm A}}=-\left[|r_{\rm B}|-\frac{m(|r_{\rm B}|)}
{|r_{\rm B}|}\right],
\label{cond1}
\end{equation}
from equation (\ref{lsq}).

The amplification factor is written by
\begin{equation}
\mu(r)=\left| \left[ 1-\frac{m(r)}{r^2} \right] \left[ 1-\frac{d}{dr}
\frac{m(r)}{r} \right] \right|^{-1}.
\end{equation}
The observed flux density ratio, $f$, should satisfy
\begin{equation}
f=\frac{\mu(r_{\rm A})}{\mu(r_{\rm B})}.
\label{cond2}
\end{equation}

From equations (\ref{cond1}) and (\ref{cond2}),
we can uniquely determine the parameters $r_{\rm s}$ and $\kappa_0$.
The characteristic overdensity, $\delta_{\rm c}$, can be obtained from these
parameters.
If the two parameters, $\delta_{\rm c}$ and $r_{\rm s}$, are determined,
the mass of the halo can also be obtained.
However, the total mass of the NFW profile is divergent at $r=\infty$.
In NFW96, $r_{200}$, which determines the mass
of the halo, was defined as $M_{200}=200 \rho_{\rm cr} (4 \pi/3) r_{200}^3$.
Namely, the mean density of the halo within $r_{200}$ is $200 \rho_{\rm cr}$.
The characteristic overdensity for the NFW profile is analytically expressed as
\begin{equation}
\delta_{\rm c}=\frac{200}{3} \frac{c^3}{\ln{(1+c)}-c/(1+c)},
\label{cNFW}
\end{equation}
where $c=r_{200}/r_{\rm s}$.

For the softened isothermal profile,
we conform the method of mass estimation to the method in NFW96.
We obtain
\begin{equation}
\delta_{\rm c}=\frac{200}{3} \frac{c^3}{c-\tan^{-1}{c}},
\label{cSIS}
\end{equation}
in a similar way.
Both $\delta_{\rm c}$'s in equations (\ref{cNFW}) and (\ref{cSIS}) are
monotone increasing functions of $c$.
We will discuss the parameter $c$,
instead of $\delta_{\rm c}$.
Since $M_{200}$ is nothing but a criterion in each profile,
we do not directly compare the masses for the two profiles.
This circumstance also holds for the other parameters also.

In NFW96, the characteristic overdensity
of the halos with masses $M_{200}$ was estimated to range
from those of dwarf galaxy halos to those of rich galaxy clusters.
The results are summarized in table 1 in NFW96.
We compare our results with those of NFW96.
In addition, we calculate the time delay,
the image position of the undetected third
image C and its flux, using the determined parameters $r_{\rm s}$ and $\delta_{\rm c}$.

\section{Results}

\subsection{Samples}

As a result of our simple assumptions for lensing galaxies,
the number of usable lens systems is limited in our estimation.
Few 2-image lens systems have been discovered.
In addition, the lensing galaxies of the lens systems have not all been found.
We exclude the double quasar 0957+561 from our samples
because a cluster of galaxies apparently contributes to the wide
separation between the images of this system
(Young et al. 1981).
The observed parameters of our samples
are listed in table 1.

We adopt the data of flux density ratio in the reddest or longest wave band
in order to minimize the influence of dust extinction.
The flux-density ratios are less well determined than the image positions,
because of the source variability, microlensing, extinction, and so on.
To account for this, we broadened the flux error bars to $20\%$
in a similar method to that of Keeton and Kochanek (1997).
The lens and image positions are not exactly on a straight line
in the observational data of the 2-image lens systems.
Generally, the observational error of a lens position is larger than one of
the image positions.
Therefore, in order to adopt our method,
we force the lens position to be placed on the nearest point
to the observed lens position on a straight line joining the two images.
We obtained the image positions in a one-dimensional form,
and list them in table 1.
The distances from the observed lens positions to the moved lens positions
are $0''.002$, $0''.035$, $0''.121$ and $0''.08$
for B1600+434, B1030+074, SBS 1520+530 and HE 1104$-$1805, respectively.
Some of them are larger than the observational errors of the lens positions.
This fact indicates that we need external shears in our analysis of these systems.

When an external shear is weak,
it slightly moves the image positions,
and breaks the alignment between the images and lens.
We can assume that the error of the image positions
in the spherical approximation, due to the external shear,
is on the same order of the lens-position shift
in the weak external shear limit.
In this case, the errors of the obtained parameters due to the error of
the image positions are smaller than those due to the flux error,
as we discuss in the next subsection.
Thus, when estimating the order of the parameters of a galaxy profile,
the effect of external shear is negligible.
We neglect the errors of image positions, which is equivalent to neglecting
the external shear as a first-order approximation.
In this method, the fitness of the statistic cannot be evaluated.
Alternatively, taking into account the flux error,
we estimate the uncertainties in the results.

The lensing galaxy of B1600+434 is an edge-on spiral
(Jaunsen, Hjorth 1997),
and that of B1030+074 has a substructure (Xanthopoulos et al. 1998).
We, however, neglect these structures in our calculation.

\subsection{Mass Profile of Galaxies}

The results of our calculation are summarized in table 2.
The errors of the results in table 2 are derived from the
flux error bar.
We neglect the errors of the image positions due to external shear.
The shift of the lens position in SBS 1520+530 is the largest one at
$0''.121$.
The errors of the image and lens positions
in this method of analysis are on the same order as the shift of the lens position.
We force the lens position in SBS 1520+530 to shift $0''.1$ from the position
given in table 1 toward images A or B,
and calculate the parameters of the lens galaxy for the NFW profile.
Next, we move the both positions of images A and B, in table 1, $0.05$ arcsec
(a total of $0''.1$) toward or away from the lens position,
and calculate the parameters.
In these cases, the maximum differences from the results in table 2a
in the scale radius and the characteristic overdensity
are $0.33$ kpc and $0.20 \times 10^7$, respectively.
These are smaller than ones due to the flux error in table 2a.
The effects of the errors due to the image positions of the other lens systems
are much smaller, and negligible.
This situation is the same for the isothermal profile.
Thus, we can neglect the errors of the image positions in this method
of analysis.

HE 1104$-$1805 has no solution for the NFW profile in equations (10) and (12).
This means that our methods cannot explain this lens system.
The system is an extraordinary system.
The mass of the lensing object is
large, because of the high redshift of the lensed images
and the wide angular separation between the images.
The redshift of the lensing object is very high
at $z=1.6616$ (Courbin et al. 1998).
A reason for the absence of a solution
might be that the system is influenced by a cluster of galaxies
or external shear.
This possibility remains to be proved
because no obvious overdensity of galaxies is detected
in the immediate region surrounding the lens (Courbin et al. 1998).
This system has a solution for the softened isothermal profile.
However, the predicted flux of the undetected third image C
is brighter than that of image B.
Therefore, our simple analysis would be undesirable for HE 1104$-$1805.
HE 1104$-$1805 needs some perturbations.

On the other hand,
the softened isothermal profile can not explain the image positions
and flux density ratio of the lens system B1030+074,
because of the extremely high flux ratio.
The flux-density ratio is one of the highest for
any known lensed system (Xanthopoulos et al. 1998).
However, the NFW profile can explain this lens system.
For B1030+074, the NFW profile is more favorable than the isothermal profile,
unless one takes into account the substructure of the lens galaxy.

The two profiles of the mass distribution have their merits and demerits
for each observation of gravitational lens system.
We cannot determine which profile is superior
from only the gravitational lens.
As a result, we conclude that the reproducibilities of image positions and flux ratios
are comparable for the two profiles.

In figure 1, we plot the dimensionless parameter $c$ against
the scale radius $r_{\rm s}$ for the NFW profile
in table 2a (filled square).
The results of numerical experiments
of $M_{200} < 10^{14} M_{\odot}$ in NFW96 are also plotted (open square).
Using $\chi^2$ minimization, they are empirically approximated by
\begin{equation}
c=37.4 \left( \frac{r_{\rm s}}{1~{\rm kpc}} \right)^{-0.3},
\end{equation}
for $M_{200} < 10^{14} M_{\odot}$.
This equation is a nothing but a gauge to make easier
to compare our results with those of the simulations.
Apparently, the values of parameter $c$ of the three lensing galaxies
are larger than the results of NFW96, and
the scale radii, $r_{\rm s}$, of the lensing galaxies
are smaller than those predicted by NFW96.
The large value of $c$ means a large characteristic overdensity.
The largest value of $c$ in figure 1 is derived from B1030+074,
the lens system of the highest flux density ratio.
We also plot $c$ against the mass
of the halo, $M_{200}$, in figure 2.
The relation of these two parameters in NFW96
is approximately expressed as
\begin{equation}
c=56.2-3.37 \log{\left( \frac{M_{200}}{M_{\odot}} \right)}
\end{equation}
for $M_{200} < 10^{14} M_{\odot}$.
Figure 2 is equivalent to figure 1,
because the NFW profile is a 2-parameter model of the matter distribution.
figure 2, however, clearly indicates that the values of $c$
are much larger than those of NFW96 for the same value of $M_{200}$.
We have compared the parameters derived from the simulations at $z=0$
with those of the high redshift lensing galaxies.
Since the central concentration of galaxies becomes larger, in general,
as galaxies evolve,
the high-redshift lensing galaxies should have smaller values of $c$
than the predicted value at $z=0$, as is in NFW96 or NFW97.
Therefore, the result, that $c$ of the lensing galaxies is too large,
is unchanged.

For the softened isothermal profile,
a diagram for $c$ versus $r_{\rm s}$ in table 2b is shown in figure 3.
The lowest value of $c$ in figure 3 is the estimate for HE 1104$-$1805.
The core radius is very large.
We also plot for the galaxy models of the softened isothermal profile
described in Jaroszy\'nski (1992).
These models have often been used to study light propagation
in an inhomogeneous universe (Jaroszy\'nski 1992; Premadi et al. 1998).
The characteristic overdensity in the models of Jaroszy\'nski is given by
\begin{equation}
\delta_{\rm c}=\frac{v^2}{4 \pi \rho_{\rm cr} r_{\rm s}^2 G},
\end{equation}
where $v$ is the velocity dispersion.
The velocity dispersions are given by
\begin{equation}
v=
\left\{ \begin{array}{ll}
390 \left( \displaystyle{\frac{L}{L_{*}}} \right)^{1/4} {\rm km~s^{-1}}, &
\mbox{(ellipticals)},\\
357 \left( \displaystyle{\frac{L}{L_{*}}} \right)^{1/4} {\rm km~s^{-1}}, &
\mbox{(S0s)},\\
190 \left( \displaystyle{\frac{L}{L_{*}}} \right)^{0.381} {\rm km~s^{-1}}, &
\mbox{(spirals),} 
\end{array} \right. 
\label{velo}
\end{equation}
where $L$ and $L_{*}$ are the luminosity and characteristic luminosity,
respectively.
The Faber--Jackson relation (Faber, Jackson 1976) and the Tully--Fisher relation
(Tully, Fisher 1977) were used for early-type and spiral galaxies, respectively.
The dependence of the scale radius on the luminosity
has been studied by Lauer (1985), Kormendy (1987), and Fukugita and Turner (1991).
Their studies are based on surface photometry profiles,
H {\small I} rotation curves, or the virial theorem which are
independent of gravitational lensing.
In conformity to these studies, the models of Jaroszy\'nski adopt
\begin{equation}
r_{\rm s}=
\left\{ \begin{array}{ll}
200 \left( \displaystyle{\frac{L}{L_{*}}} \right) \mbox{pc},&
\mbox{(ellipticals and S0s)},\\
2 \left( \displaystyle{\frac{L}{L_{*}}} \right) \mbox{kpc}, &
\mbox{(spirals).} 
\end{array} \right. 
\label{rs-l}
\end{equation}
Eliminating $L/L_{*}$ by equations (\ref{velo}) and
(\ref{rs-l}), the $c$--$r_{\rm s}$ relation can be plotted.

In figure 3, all data points of the gravitational lenses,
including HE 1104$-$1805, are obviously
close to the lines for the models of Jaroszy\'nski for early-type galaxies.
Although the lensing galaxy of B1600+434 is a spiral,
the models for early-type galaxies are favorable.
The value of $c$ given by the model for spirals
is smaller than those of the lensing galaxies.
$r_{\rm s}$ of HE 1104$-$1805 is too large.
This system needs some perturbations as we have mentioned above.
However, the orders of $c$ in the other two systems are available.
On balance, there is no serious problem using the isothermal sphere
as a first approximation of a cosmological tool.

We cannot directly compare the consistencies of the NFW and the softened isothermal profile.
However, the difference between the results of NFW96 and ours is conspicuous,
while the models of Jaroszy\'nski well account for the parameters of
the lensing galaxies.
As we shown above, the softened isothermal sphere
as a first approximation of a cosmological tool
has no serious problem
in an analysis of light propagation in a universe.
We cannot use the NFW profile, without stars,
and need more complicated assumptions in such analysis.

The predicted time delays of B1600+434 and SBS 1520+530 for the NFW profile
in table 2a
are about $10\%$ larger than those for the isothermal profile in table 2b.
This different time delay for the two profiles
will sufficiently affect the estimate of the Hubble constant.

\section{Conclusions and discussion}

\indent

We applied the NFW and the softened isothermal profile
to lensing galaxies.
We assumed that there is no contribution of stars and external shear,
and that the lens galaxies are spherically symmetric, as a first approximation.
By reproducing the image positions and their flux ratios,
we estimated the scale radii and the characteristic overdensities.
The reproducibilities of image positions and flux ratios
are comparable for the two profiles.
If we assume that the mass distribution of galaxies
is described by the NFW profile,
the scale radii are smaller, and the characteristic overdensities
are larger than those in NFW96.
If we adopt the softened isothermal profile to the lensing galaxies,
the scale radii and the central matter densities are not so
different from those in the models of Jaroszy\'nski for early-type galaxies,
which are derived from the various observational results.
The central densities of the Jaroszy\'nski model for spiral galaxies
are too small to account for the observations of the lensing galaxies,
including the spiral lens in B1600+434.
The predicted time delays for the NFW profile
are substantially larger than those for the isothermal profile.

If we uncritically accept these results,
they are more advantageous to the softened isothermal profile than to the NFW profile.
However, there are two possible and natural explanations
of our results for the NFW profile.
One is that our lens model may be overly simplified.
We assumed that the contributions of stars and gas are negligible.
A large amount of stars may influence the image deflection.
We cannot simply adopt the NFW profile, which does not include stars,
to probe the cosmological
parameters or the light propagation in an inhomogeneous universe.
We need more complicated models with stars in their analyses.
We also assumed that the external shear is weak.
If the external shear is strong, our method cannot be applied.

Another explanation is that the results in NFW96 for galaxies are wrong
due to softening and small numbers of particles.
In this case, much higher resolution is required in {\it N}-body simulations
in order to describe the parameters of galaxies.
For the present, we cannot decide which explanation is more accurate.
Therefore, we need much more sophisticated lens models
which include the contribution of stars or external shear,
which we plan to develop in the future.
Anyhow, the NFW profile for galaxies, which does not include stars,
cannot be applied as a first approximation of galaxy model at this stage.
The isothermal sphere as a first approximation of a galaxy model
has no serious problem.

We discuss other possible explanations for our results below.
In NFW96, the smallest gravitational
softening is 1.5 kpc.
This size could be too large to estimate the scale radius
of galaxies.
However, the small gravitational softening may influence
the central profile of galaxies in {\it N}-body simulations.
The power of the density cusp at the center is
controversial because of effects of gravitational softening and
poor resolution at small scales. Fukushige and Makino (1997) showed
from simulations that as particle numbers increase 1 order of magnitude,
typical central density profiles become shallower than $\rho\propto
r^{-2}$, but steeper than $\rho\propto r^{-1}$.
~Moore et al. (1998) also indicate that
the central density profile of cold dark-matter halos
becomes steeper as the resolution
in the cosmological {\it N}-body simulations becomes higher.
The density profiles of Moore et al., or other steep profiles, need to be adopted
to the lensing galaxies.
These are our future objectives.

As mentioned in subsection 3.2,
the effects of galactic evolution cannot remedy the contradiction
of the characteristic overdensity,
because the central concentration of galaxies generally becomes larger
as the galaxies evolve.
The high-redshift lensing galaxies should have smaller values of $c$
than the predicted value.

If we assume that the initial density
fluctuation spectra, $P(k) \propto k^n$, is not restricted by $\sigma_8$,
the rms mass fluctuation in spheres of radius $8 h^{-1}$ Mpc,
we can enlarge the value of $c$ in {\it N}-body simulations.
While the standard model, $\sigma_8=0.63$, was assumed in NFW96,
a universe with various power spectra in NFW97, $n=0$, $-0.5$, $-1$, and $-1.5$,
produces larger values of $c$.
As index $n$ becomes larger, the value of $c$ becomes larger.
From figure 9 in NFW97,
universes with indices $n=0$, or $-0.5$ seem favorable for our results
on gravitational lensing.

If the cosmological constant, $\Lambda_0$, has a finite value,
we can marginally decrease the value of $c$ for the lensing galaxies.
For example, $c$'s values are $77.7$, $299$, and $68.1$ for B1600+434,
B1030+074, and SBS 1520+530, respectively,
in the flat $\Lambda_0$ universe; $H_0=75~{\rm
km\,s^{-1}\, Mpc^{-1}}$, $\Omega_0=0.25$, and $\Lambda_0=0.75$.
Here, we expressed $\Lambda_0$ in units of $3 H_0^2$.
However, judging from figure 9 in NFW97, these results do not
agree with the {\it N}-body simulations.

Of course, the statistical reliability of our estimation
is not high enough, because of the small number of samples.
As mentioned, the number of 2-image lens systems is not very large.
More sophisticated lens models
can be adopted to more complex lens systems
and we will have to increase the number of samples.

\vspace{2pc}\par

We are grateful to Nobuyoshi Makino and Kenji Tomita
for their helpful advice.
This work was supported by a Research Fellowship of the Japan Society for
the Promotion of Science.
\vfill\eject

\section*{References}
\small

\re
	Bartelmann M.\ 1996, A\&A\ 313, 697

\re
	Binney J., Tremaine S.\ 1987, Galactic Dynamics
	(Princeton University Press, Princeton) ch4

\re
	Carlberg R. G., Yee H. K. C., Ellingson E., Morris S. L., Abraham. 
	R., Gravel P., Pritchet C. J., Smecker-Hane T. et al.\ 1997, ApJ\ 485, L13

\re
	Courbin F., Lidman C., Magain P.\ 1998, A\&A\ 330, 57

\re
	Crampton D., Schechter P. L., Beuzit J.-L.\ 1998, AJ\ 115, 1383

\re
	Eke V. R., Navarro J. F., Frenk C. S.\ 1998, ApJ\ 503, 569

\re
	Evans N. W., Wilkinson M.\ 1998, MNRAS\ 296, 800

\re
	Faber S. M., Jackson R.\ 1976, ApJ\ 204, 668

\re
	Fassnacht C. D., Cohen J. G.\ 1998, AJ\ 115, 377

\re
	Fukushige T., Makino J.\ 1997, ApJ\ 477, L9

\re
	Fukugita M., Turner E. L.\ 1991, MNRAS\ 253, 99

\re
	Grogin N. A., Narayan R.\ 1996, ApJ\ 464, 92


\re
	Jaroszy\'nski M.\ 1992, MNRAS\ 255, 655

\re
	Jaunsen A. O., Hjorth J.\ 1997, A\&A\ 317, L39

\re
	Keeton C. R., Kochanek C. S.\ 1997, ApJ\ 487, 42

\re
	Kent S. M., Falco E. E.\ 1988, AJ\ 96, 1570

\re
	Koopmans L. V. E., de Bruyn A. G., Jackson N.\ 1998, MNRAS\ 295, 534

\re
	Kormendy J.\ 1987, in Dark Matter in the Universe,
	ed J.Kormendy, G.R.Knapp, IAU Symp. No.117
	(Reidel, Dordrecht) p139

\re
	Lauer T. R.\ 1985, ApJ\ 292, 104

\re
	Makino N., Asano K.\ 1999, ApJ\ 512, 9

\re
	Makino N., Sasaki S., Suto Y.\ 1998, ApJ\ 497, 555

\re
	Maoz D., Rix H.-W., Gal-Yum A., Gould A.\ 1997, ApJ\ 486, 75

\re
	Moore B., Governato F., Quinn T., Stadel J., Lake G.\ 1998, ApJ\ 499, L5

\re
	Navarro J. F., Frenk C. S., White S. D. M.\ 1995, MNRAS\ 275, 720

\re
	Navarro J. F., Frenk C. S., White S. D. M.\ 1996, ApJ\ 462, 563 (NFW96)

\re
	Navarro J. F., Frenk C. S., White S. D. M.\ 1997, ApJ\ 490, 493 (NFW97)

\re
	Premadi P., Martel H., Matzner R.\ 1998, ApJ\ 493, 10

\re
	Schneider P., Ehlers J., Falco E. E.\ 1992, Gravitational Lenses
	(Springer, Berlin) ch8

\re
	Sofue Y.\ 1999, PASJ\ 51, 445

\re
	Sofue Y,. Tomita A., Tutui Y., Honma M., Takeda Y.\ 1998, PASJ\ 50, 427

\re
	Sofue Y., Tutui Y., Honma M., Tomita A.\ 1997, AJ\ 114, 2428

\re
	Tomita K., Premadi P., Nakamura T.\ 1999,
	Prog. Theor. Phys. Suppl.\ 133, 85

\re
	Tully R. B., Fisher J. R.\ 1977, A\&A\ 54, 661

\re
	Xanthopoulos E., Browne I. W. A., King L. J., Koopmans L. V. E.,
	Jackson N. J., Marlow D. R., Patnaik A. R., Porcas R. W.,
	Wilkinson P. N.\ 1998, MNRAS\ 300, 649

\re
	Young P., Gunn J. E., Kristian J., Oke J. B., Westphal J. A.\ 1981,
	ApJ\ 241, 507

\vfill\eject
\begin{center}
{\bf Figure Captions}
\end{center}

\begin{description}
\item[Fig. 1.] Diagram for the dimensionless parameter $c$ against
the scale radius $r_{\rm s}$ for the NFW profile.
The filled squares are derived from our results in table 2a.
The error bars derived from the $20\%$ error of the flux ratio are also
plotted.
The open squares are the results of the {\it N}-body simulations
for $M_{200} < 10^{14} M_{\odot}$ in table 1 in NFW96.
The solid line is the best-fit relation between $c$ and $r_{\rm s}$
for the results of NFW96.

\item[Fig. 2.] Diagram for the dimensionless parameter $c$ against
the mass $M_{200}$ for the NFW profile.
The filled squares, the open squares, and error bars
are the same as in figure 1.
The solid line is the best-fit relation between $c$ and $M_{200}$
for the results of NFW96.

\item[Fig. 3.] Diagram for the dimensionless parameter $c$ against
the scale radius $r_{\rm s}$ for the softened isothermal profile.
The filled squares were derived from our results in table 2b.
The error bars are the same as in figure 1.
The solid line, dotted line, and short dashed line indicate
the models of Jaroszy\'nski for elliptical, S0, and spiral galaxies, respectively.

\end{description}

\newpage

\begin{table*}[h]
\begin{center}
Table~1. \hspace{4pt}List of 2-image gravitational lens systems.$*$
\end{center}
\vspace{6pt}
\begin{tabular}{lcccccc}
\hline \hline
Name & $z_{\rm l}$ & $z_{\rm s}$ & $r_{\rm A}$ & $r_{\rm B}$ & $f$ &
 References \\
&&&arcsec (kpc)&arcsec (kpc)&& \\ \hline
B1600+434 & 0.4144 & 1.589 & 1.14 (7.5)& -0.25 (1.6)& 1.25 & 1,2 \\
B1030+074 & 0.599 & 1.535 & 1.37 (10.4) & -0.16 (1.3)& 14 & 1,3 \\
SBS 1520+530 & 0.765 & 1.855 & 1.19 (9.7)& -0.38 (3.1)& 1.89 & 4 \\
HE 1104$-$1805 & 1.6616 & 2.316 & 1.14 (9.6)& -2.06 (17.4)& 3.47 & 5 \\ \hline
\end{tabular}
\vspace{6pt}\par\noindent
$*$ $z_{\rm l}$ and $z_{\rm s}$ are redshifts of the lens and the source,
respectively.
references.-(1) Fassnacht and Cohen (1998);
(2) Koopmans, de Bruyn, and Jackson (1998);
(3) Xanthopoulos et al. (1998);
(4) Crampton, Schechter, and Beuzit (1998);
(5) Courbin, Lidman, and Magain (1998).
\end{table*}

\begin{table*}[h]
\begin{center}
Table~2(a). \hspace{4pt}Results for the NFW profile.$*$
\end{center}
\vspace{6pt}
\begin{tabular}{lccccccc}
\hline \hline
Name & $r_{\rm s}$ & $\delta_{\rm c}$ & $c$ & $M_{200}$ &
$r_{\rm C}$ & $f_{\rm CB}$ & Time Delay \\
& kpc & $10^7$ && $10^{12} M_{\odot}$ & arcsec && days \\ \hline
B1600+434 & $2.85^{+0.34}_{-0.28}$ & $1.49^{+0.34}_{-0.30}$ & $92.6^{+7.2}_{-7.2}$ &
$1.08^{+0.11}_{-0.09}$ & $-0.16^{+0.01}_{-0.01}$ & $0.49^{+0.07}_{-0.06}$ & 
$59.2^{+1.8}_{-2.0}$ \\
B1030+074 & $0.58^{+0.12}_{-0.10}$ & $69.9^{+36.0}_{-23.7}$ & $372^{+59}_{-51}$ &
$0.593^{+0.066}_{-0.055}$ & $-0.04^{+0.01}_{-0.01}$ & $0.07^{+0.05}_{-0.03}$ & 
$207.2^{+2.3}_{-2.7}$ \\
SBS 1520+530 & $3.85^{+1.37}_{-0.90}$ & $1.31^{+0.95}_{-0.59}$ & $88.3^{+19.5}_{-17.6}$ &
$2.31^{+0.65}_{-0.41}$ & $-0.11^{+0.02}_{-0.03}$ & $0.13^{+0.07}_{-0.05}$ & 
$141.7_{-12.6}^{+10.4}$ \\ 
HE 1104-1805 &\multicolumn{7}{c}{no solution} \\ \hline
\end{tabular}
\vspace{6pt}\par\noindent
$*$ $r_{\rm C}$ is the image position of the undetected third image.
$f_{\rm CB}$ is the flux density ratio of image C to B.
~The time delays of the image B lagging behind A are also listed.
\end{table*}

\begin{table*}[h]
\begin{center}
Table~2(b). \hspace{4pt}Results for the softened isothermal profile.
\end{center}
\vspace{6pt}
\begin{tabular}{lccccccc}
\hline \hline
Name & $r_s$ & $\delta_{\rm c}$ & $c$ & $M_{200}$ &
$r_{\rm C}$ & $f_{\rm CB}$ & Time Delay \\
& pc & $10^7$ && $10^{13} M_{\odot}$ & arcsec && days \\ \hline
B1600+434 & $519^{+50}_{-47}$ & $9.77^{+1.87}_{-1.49}$ & $1210^{+111}_{-96}$ &
$1.46^{+0.04}_{-0.03}$ & $-0.17^{+0.01}_{-0.02}$ & $0.50^{+0.08}_{-0.08}$ & 
$49.3^{+0.7}_{-0.7}$ \\
B1030+074 & \multicolumn{7}{c}{no solution} \\
SBS 1520+530 & $715^{+254}_{-232}$ & $8.25^{+8.99}_{-3.52}$ & $1112^{+496}_{-270}$ &
$2.96^{+0.23}_{-0.20}$ & $-0.12^{+0.04}_{-0.04}$ & $0.15^{+0.10}_{-0.08}$ & 
$117.5^{+5.4}_{-5.8}$ \\ 
HE 1104-1805 & $19200^{+4400}_{-3900}$ & $0.179^{+0.057}_{-0.038}$ & $163^{+24}_{-19}$ &
$183^{+52}_{-41}$ & $0.77^{+0.04}_{-0.06}$ & $2.49^{+0.68}_{-0.68}$ & 
$-242.2_{-88.8}^{+65.6}$\\ \hline
\end{tabular}
\end{table*}

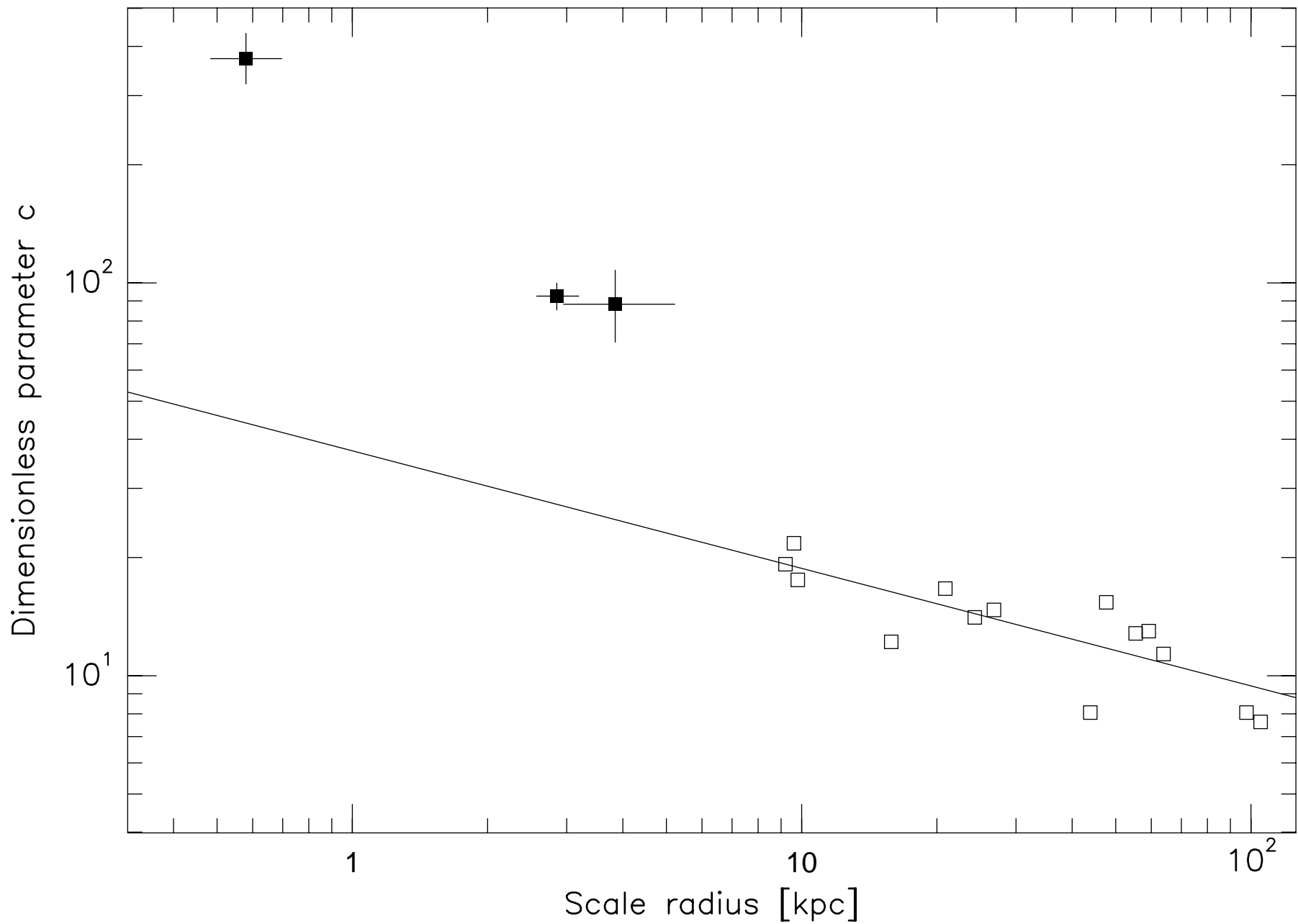
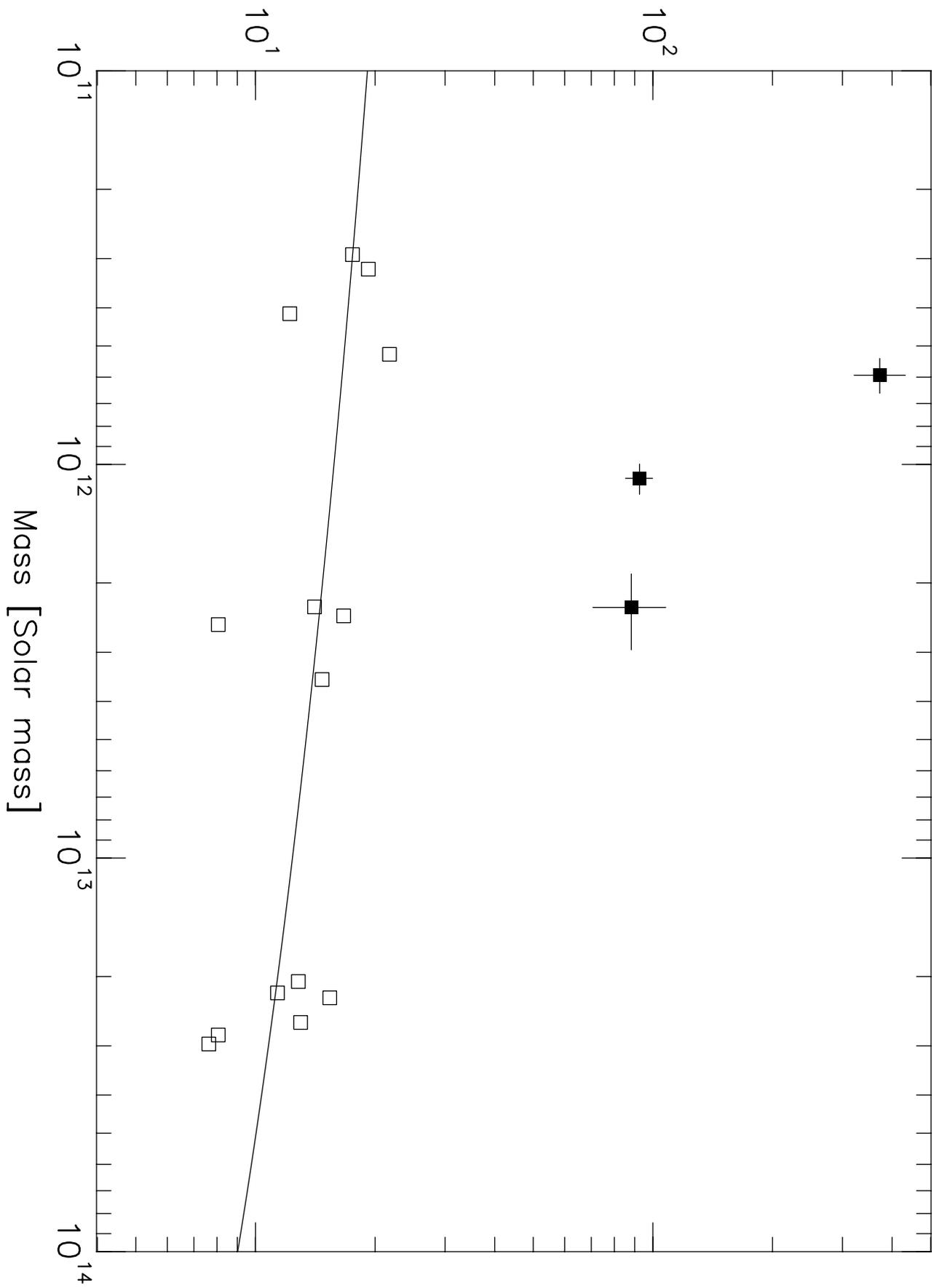
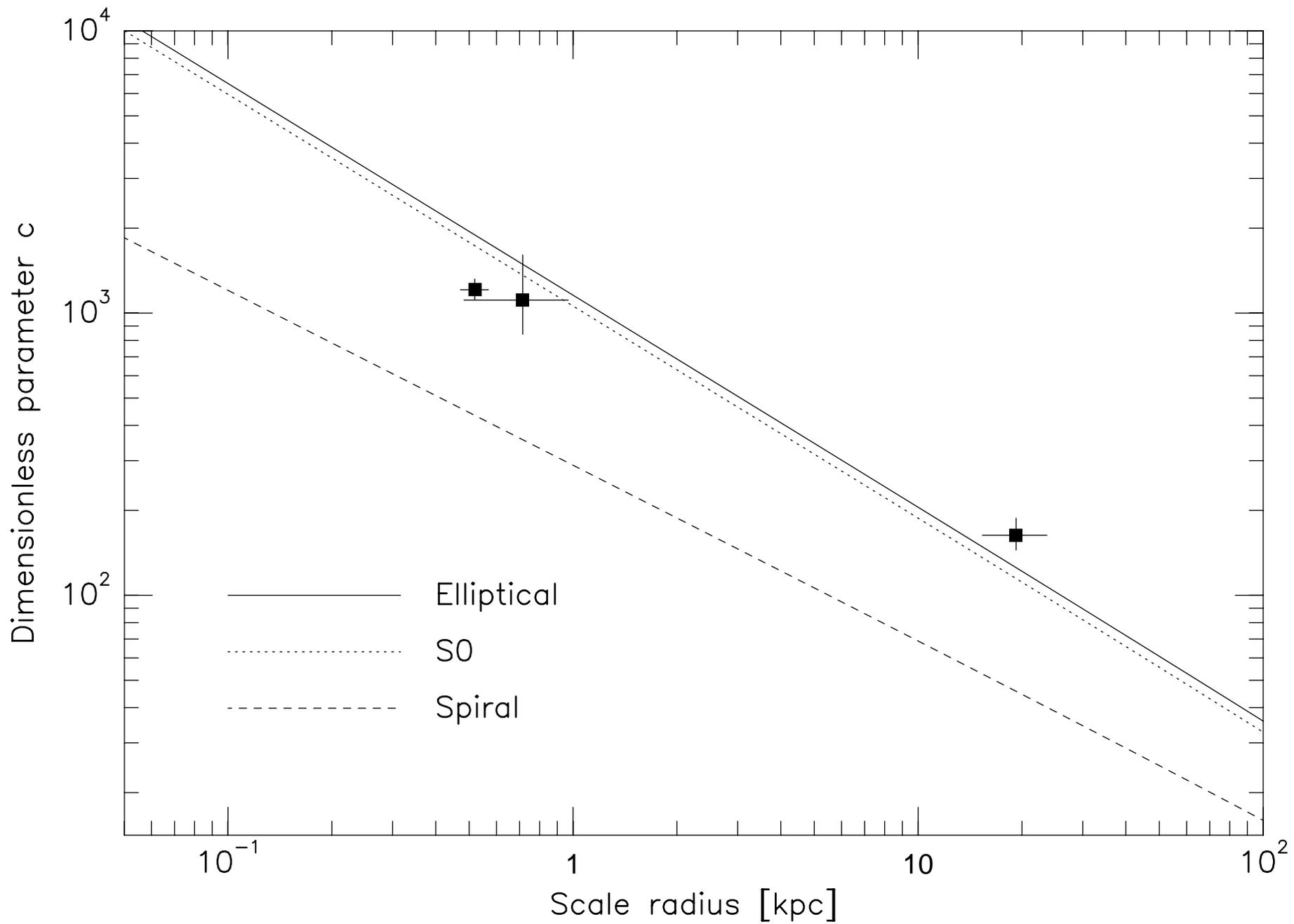
\end{document}